\documentclass[10pt]{article}

\RequirePackage{graphics}

\title{Contextual Constraint Modeling in \\
          Grid Application Workflows}
\author{Greg Graham, Anzar Afaq, David Evans, \\
        Gerald Guglielmo,  Eric Wicklund\\
{\em  Fermi National Accelerator Laboratory, Batavia, IL, 60510, USA}\\ \\
Peter Love\\
{\em Department of Physics, Lancaster University, Lancaster, UK} 
\\ \\
Submitted to \\
Concurrency and Computation: Practice and Experience \\
Special Edition for GGF10 Workflows Section
}

\begin{document}

\maketitle
\begin{abstract}
This paper introduces a new 
mechanism for specifying constraints in distributed 
workflows.  By introducing constraints in a contextual form, it is shown how 
different people and groups within collaborative communities can cooperatively 
constrain workflows.  A comparison with existing state-of-the-art workflow 
systems is made.
These ideas are explored in practice
with an illustrative example from High Energy Physics.
\end{abstract}

\section{Introduction}

The Grid is emerging as a specialized distributed computation standard of 
unprecedented power and scope, promising to turn commodity networks and 
computers into commodity computation.  The Grid concept has already been 
proven useful for science in many applications \cite{bib:foster04}\cite{bib:berman03}.  Substantial 
infrastructure already exists \cite{bib:grid2003} or is being planned \cite{bib:egee}\cite{bib:osg}.  
Data processing 
on the Grid ranges from tightly coupled kinds 
of problems to loosely coupled (the so-called 
"embarrassingly parallel") ones.  
Computation involving the use of MPI \cite{bib:mpi} or some 
other parallel processing standard within a single application 
is an example of the former kind. 
The High Energy Physics (HEP) problem domain is characterized by the 
latter: parallel independent filtering or 
data processing applications with large data flows 
between them. Requirements placed on large processing projects 
are often coordinated among many interests and groups 
within a Virtual Organization (VO).  In the case of 
workflows 
containing a moderate number of application steps, it becomes a daunting task 
to check that all of the input parameters and installed software 
configurations conform to 
decisions made at the collaboration level.  It is useful therefore to have a 
mechanism with which it is able to specify constraints on the parameters of the 
individual 
workflow steps that bring them into line with collaborative decisions 
coherently across the entire workflow, possibly even dynamically as decisions 
or discoveries are being made.  In the paper, it will be shown that:
\begin{itemize}
\item Collections of constraints can be gathered into documents called 
contexts 
that function as operators on existing workflow graphs.  An algebra of 
contexts supporting factoring and 
composition can help different groups within a VO work 
together though constraint sharing.  Decomposition of contexts can allow for 
variance of constraints simultaneously across several different categories.
\item Constraint expressions and contexts form an interesting and hitherto 
largely unexplored area of data provenance.  Knowledge of the constraints 
implies that it is possible not only to know the values of application input 
parameters, but also why they were set in particular ways.
\end{itemize}
It can be assumed that such constraints can be distributed within 
the stack of a single running application using MPI.  However, the techniques 
developed here find fruitful application in the organizational 
aspects and in the aspects of sharing collaborative decisions about 
constraints in the multi-application domain.  
A partial implementation of these ideas already exists in  
workflow building tools \cite{bib:runjob}\cite{bib:ggraham03} developed 
for the Compact Muon Solenoid 
(CMS) \cite{bib:cms} experiment, a HEP experiment based at the European 
Center for Nuclear Research (CERN) \cite{bib:cern} in Geneva, Switzerland.

In the following, we will focus mainly on semantic constraints and not 
constraints on physical resources, synchronization, nor parallelization.  
Many traditional workflow specification schemes such as DAGMan \cite{bib:dagman} and 
constraint languages such as ClassAds \cite{bib:classad} already address these concerns.  
In sections \ref{sc:constraints} and \ref{sc:context}
it will be shown how multigraphs, including new arrow types called metadata 
flows, can be used to express constraints.  It will also be shown how 
these constraints can be expressed in cooperating context documents. 
And a general 
procedure for reducing multigraphs into fully constrained workflow 
descriptions suitable for execution by a workflow manager such as DAGMan 
will be outlined. Section \ref{sc:applications} will illustrate the concepts
introduced in previous sections with an illustrative example from 
High Energy Physics.  Finally, the 
discussion will be wrapped up and comparisons 
with existing workflow systems will be made in the conclusion.

\section{Objects and Operations in a Workflow Constraint Specification}
\label{sc:constraints}

Workflow specifications are often expressed as directed graphs.  In many 
cases where the workflow consists of pure filtering and/or simple 
single-pass processing, these graphs are directed acyclic graphs (DAGs).  
Let $G=(N,A)$ be a general unconstrained workflow graph.  Each node $N$ in $G$ corresponds to an 
application and the set of arrows $A$ corresponds to a partial sequencing of 
the nodes, usually generated by real data flow relationships\footnote{Alternatively, the nodes may 
correspond to data products and the arrows may correspond to data transformations,
as in the Chimera Virtual Data System \cite{bib:chimera}.  
These two pictures are equivalent.}.  Each of these nodes may 
have attributes specifying input parameters to or conditions on the 
corresponding application.  In order to express constraints on $G$, one may
add both nodes and arrows.  The resulting data structure is a 
multigraph\footnote{A graph comprises some set $N$ of nodes and a set $A$ of arrows 
such that for any two nodes $N$ and $M$ in $N$, $N$ and $M$ can have at most 
one arrow $a$ in $A$ between them.  In a multigraph, the restriction on the 
number of arrows is dropped.}.   
\begin{figure}[hbtp]
  \begin{center}
    \resizebox{14cm}{!}{\includegraphics{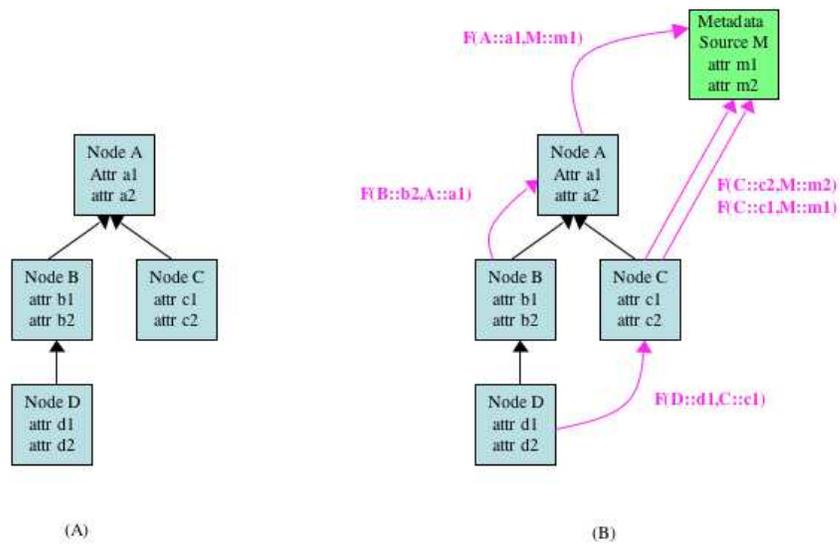}}
    \caption{(A) A simple four application acyclic workflow graph.  In order to execute the 
graph, all node attributes must be set.  (B) This simple graph is augmented with metadata 
flows (pink) specifying various constraints on the node attributes.  In addition a metadata
source node has been added.  The constraint augmented workflow is now a multigraph.}
    \label{fig:ofig1}
  \end{center}
\end{figure}
The extra arrows correspond to constraint relationships 
between specific node attributes.  These extra arrows will be called metadata 
flows, and the set of all metadata flow arrows constraining a graph $G$ will be 
denoted by $F$.  In addition to metadata flows, special nodes may be added whose 
only purpose is to serve as sources or sinks for metadata flows.  These extra 
nodes are called metadata terminals, and the set of these will be 
denoted $M$. This is illustrated in figure \ref{fig:ofig1}.  
An example of a metadata source terminal is a null workflow node 
that holds a query result from some catalog. An example of a metadata 
sink is a node that may consume metadata merely to record it, such as a 
tracking system or provenance recorder. In general, metadata sources and sinks may be 
replaced by a single source and a single sink node, but we may also allow for many.  
The multigraph containing $G$ and its constraints expressed by $F$ 
and $M$ will be denoted $M_G$. This is the constrained and 
unreduced workflow graph.  Several types of arrows are 
evident in $M_G$ including conventional workflow sequencing arrows 
and metadata flows.  
For a summary of all of the symbols being introduced, please refer to table \ref{tb:tb1}. 

Operations on the unreduced constrained graph consist of reduction 
operations that satisfy the metadata flows and thereby gradually 
reduce $M_G$ to a 
constrained and reduced $G'$, or a workflow graph with zero metadata flows.  
Let $F(I.i,J)$ denote a metadata flow, 
where $I$ is a node in $N$ and $J$ is a node in $N+M$, 
and $i$ is an attribute in $I$.  The first argument 
is the target of the constraint while the second argument is the domain.  A 
reduction $R_F$ over $F$ is an operation that replaces the value of attribute 
$I.i$ by some value which satisfies the constraint computed from the 
domain $J$.  For example, the simplest such operation is just the assignment 
reduction $=_F$.  The constraint $I.i =_F J.j$ on $I.i$ is that it be equal to $J.j$ 
where $j$ is some attribute in $J$.\footnote{Or, more generally, $j$ can be a simple 
expression involving attributes of $J$.}  Categorically, the unifying character of 
this picture can be seen by considering that general constraint reductions 
targeting an attribute value in some workflow node are equivalent to constant 
assignment reductions emanating from a single metadata terminal node \cite{bib:pierce91}.  

\begin{table}
\centering{
\begin{tabular}{|c|c|c|c|}
\hline
Graph & name & Nodes & Arrows \\
\hline
$G$ & Workflow Graph & $N$ & $A$ \\
$F$ & Metadata Flows & - & $F$ \\
\hline
$M$ & Metadata Terminals & $M$ & $-$ \\
\hline
$C$ & Context & $M$ & $F$ \\
\hline
$M_G$ & Constrained Unreduced Workflow & $M+N$ & $A+F$ \\
\hline
$M_G'$ & Metadata Flow Subgraph & $M+N$ & $F$ \\
\hline
$G'$ & Constrained and Fully Reduced & $N$ & $A$ \\
\hline
\end{tabular}
\caption{The above table summarizes the different graph entities introduced in 
in section \ref{sc:constraints}.  (Context will be defined in section \ref{sc:context}.)}}
\label{tb:tb1}
\end{table}

Let $M_G'$ be the metadata flow subgraph of $M_G$ such that $M_G'$ contains all of the 
nodes of $M_G$ but only the arrows in $F$.  It should be noted that this subgraph 
should be acyclic so that at least one serialization of the 
metadata flows $F$ exists. Otherwise the constraint model is 
undefined.   Metadata flows may exist to 
carry metadata to a finite number of graph nodes in a possibly cyclic $G$,  
and each node in $G$ has a finite number of attributes. Thus it is always 
possible to find a finite spanning tree if there are no more than one metadata 
flow per target attribute.

Reduction always results in the removal of a single arrow from $F$ and possibly 
the alteration of at most one attribute in the target node of the flow.  
The order of reduction is determined by a partial ordering of the nodes in 
$M_G'$.  Another 
possibility is to check and raise an exception in case a boolean constraint evaluates
to false.  Reduction can continue until $M_G$ has been transformed into $G'$.  
Nonetheless, there are many such 
reduction partial orders.  Some optimization may be gained by grouping some 
reduction operations together if it is known in advance that they can be reduced 
together.  

\section{Constraints and Contexts}
\label{sc:context}

It is useful at this point to introduce the 
context $C$, consisting of the metadata terminals alone and the metadata 
flows.  By partitioning $M_G$ into a context part $C$ and an application workflow 
part $G$, we gain the possibility that a single set of constraints agreed 
upon by some large organization can be applied to multiple application 
workflow graphs, and that a given application workflow graph can be run 
in a variety of contexts.  Furthermore, we gain the possibility of factoring 
the context $C$ itself into parts that are of interest to specific individuals or groups. 

\begin{figure}[hbtp]
  \begin{center}
    \resizebox{12cm}{!}{\includegraphics{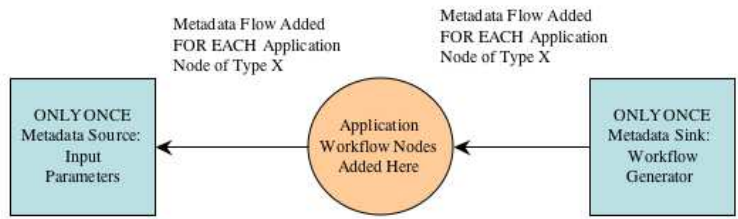}}
    \caption{An illustration of a context document.  
The application workflow node type (added in the center
center) is used to lookup rules for adding metadata flows and/or metadata 
terminal nodes.  Generally, flows are added with "for-every" semantics and 
nodes are added with "only-once" semantics.}
    \label{fig:ofig2}
  \end{center}
\end{figure}

Given the wide variety of workflow graphs and contexts possible, 
it is a non-trivial task to design an algebra by which these sets can be 
combined in a simple way.  A further problem is that $C$ by itself is not 
even a well-defined graph because some of the arrows in $C$ point to nodes that 
are not in $C$.  The approach taken here to deal with these problems is to 
assign types to the graph nodes in the application workflow graph.  By 
surveying the set of all possible application nodes in an organization, 
it is possible to come up with a context document $D_C$ that, rather 
than being a graph subset, is a collection of rules for how to apply 
metadata terminals and metadata flows in a real application workflow 
graph as workflow nodes are added into the context document.  The rules 
are indexed by node type and may be applied under one of two kinds of 
semantics: "only-once" semantics or "for-each" semantics.  For metadata 
terminals, the only-once semantics are generally used.  For metadata flows, 
the for-each semantics are generally used.  This is illustrated in figure \ref{fig:ofig2}.
This approach works for a large number of cases.  

An algebra for combining different context documents is being 
developed.  This is somewhat more difficult to do in complete generality not 
only since the documents are not graphs, but also because rules 
must be developed to handle metadata collisions when metadata flows share 
the same target. In general we would like to avoid that the final metadata 
flow that gets applied depends upon 
the order in which the context documents are processed.  Even in the absence 
of a full collision resolution algorithm, there is 
still a large class of problems for which the metadata flows do not collide 
and are yet useful.  There is also another set of problems for which the 
"shadowing" behavior of overlapping metadata flows is actually desirable, such as for 
simple replacement of site dependent variables with a site independent default.

\section{Applications}
\label{sc:applications}

Many of the ideas presented in this paper have been implemented in the
area of data processing for HEP. Two related software
toolkits have been developed for that application and will be discussed here:
\begin{itemize}
\item The RunJob Project at Fermilab \cite{bib:runjob} provides basic entities 
to help define, configure, and execute a workflow.   RunJob has metadata
terminals, metadata flows, a reduction algorithm, and context documents.
\item MCRunjob is a package to create Monte Carlo simulation jobs for the CMS 
 \cite{bib:cms} experiment.  MCRunjob is based upon the software provided
by the RunJob Project.
\end{itemize}
In the following, the problem of offline data processing and analysis
in HEP will be described in very broad terms.  Then the 
basic entities provided by the RunJob Project will be described, and a 
concrete example of the contextual application of constraints will be 
described.  

\subsection{High Energy Physics Data Analysis}

The expected volume of data from the Large Hadron Collider (LHC) experiments 
at CERN \cite{bib:cms}\cite{bib:atlas} is expected
to be very large, on the order of several petabytes of data per year.  
Distributed processing of this data is expected to be the norm instead 
of the exception, both because of the data volume and because the 
collaborators on these experiments are expected to be able to make 
significant contributions to LHC analysis without traveling to CERN for 
extended periods of time. New tools and frameworks must be brought to bear 
to help organize the resources towards the successful processing and 
analysis of data under these circumstances, and this in fact generates
the interest of HEP in Grid technology.

The successful analysis of HEP data involves the production 
and analysis of large amounts of simulated "Monte Carlo" data also in order to 
understand detector responses and biases that could affect a measurement.  
The amount of Monte Carlo data produced will be of the same order of size 
as the actual data collected.  The production of Monte Carlo usually 
involves a sequence of application programs.  The following are just three broad examples.
\begin{itemize}
\item Generator program:  This program randomly 
generates a pure physics event taking into account theoretical probabilities
for the production of subatomic particles and fast decay products associated  
with a collision event. 
\item Simulator program:  This program steps each particle created by the 
generator through a precise model of the collider detector, and calculates
the amount of energy deposited in each defined detector volume.  An 
LHC generated event may contain dozens of individual particles to track, 
but the collider detector may contain hundreds of thousand to millions of 
individually modeled detector volumes. 
\item Digitizer program:  An "active" detector volume is one which is 
instrumented in the corresponding real detector. For each active detector 
volume, this program calculates a digitized electronic signal taking into 
account what is known at the time about the electronics.  
\end{itemize}
The perennial problem in the area of distributed Monte Carlo production is  to 
constrain the applications so that the data produced at one site is of the same quality as data 
produced at any other site.  

User analysis is generally approached from the standpoint of how to gain 
individual access to distributed resources and how to move user written analysis 
software to the 
job execution site.  However, here we are more concerned with the organizational 
principles of coordinating the processing data across different user jobs.  One 
possible way to think about this problem is as a constraint.  Say users
belong to more than one physics analysis group.  Each application that 
the user submits may have dozens to hundreds of parameters to set, and 
the user will not be able in general to set them all by hand.  Often, it 
is the physics group that provides help in setting all of the "expert" parameters.  
In other words, the physics group provides a context within which the 
user must analyze data.  This is important especially if the user belongs to more 
than one physics group.  These might include the following.
\begin{itemize} 
\item  Detector parameters that only experts know how to set.  
For example, thresholds on the cells in the electromagnetic 
calorimeter.  These parameters control whether or not an individual detector 
component will write data into the output stream (i.e.- only if the simulated 
signal is above threshold)  
\item  Choice of algorithm to use for the reconstruction of a physics object.  For example, 
jet clustering, and parameters such as a cone size for that algorithm.  
Such parameters are often set by algorithm
specialists and maybe not often set by individual users unless they 
are doing a study.  
\item  Choice of input dataset.  Modern data management systems such as the 
CMS data management system will have the ability to select datasets by 
a query on the physics metadata.  The queries may differ from group to group.
\item User written software.  This could contain special user parameters or 
the specification of special libraries.
\end{itemize}
In many cases, it is important when comparing two different analyses that such parameters
as outlined above are known and controlled for.
Other site dependent parameters best set by an administrator include the following.
\begin{itemize}
\item locating the input data
\item correct placement of the output data.
\item locating of the experiment application software and libraries
\item accessing local batch or Grid resources 
\item configuring application parameters governing simulation
\end{itemize}

\subsection{The RunJob Project}

\begin{figure}[hbtp]
  \begin{center}
    \resizebox{10cm}{!}{\includegraphics{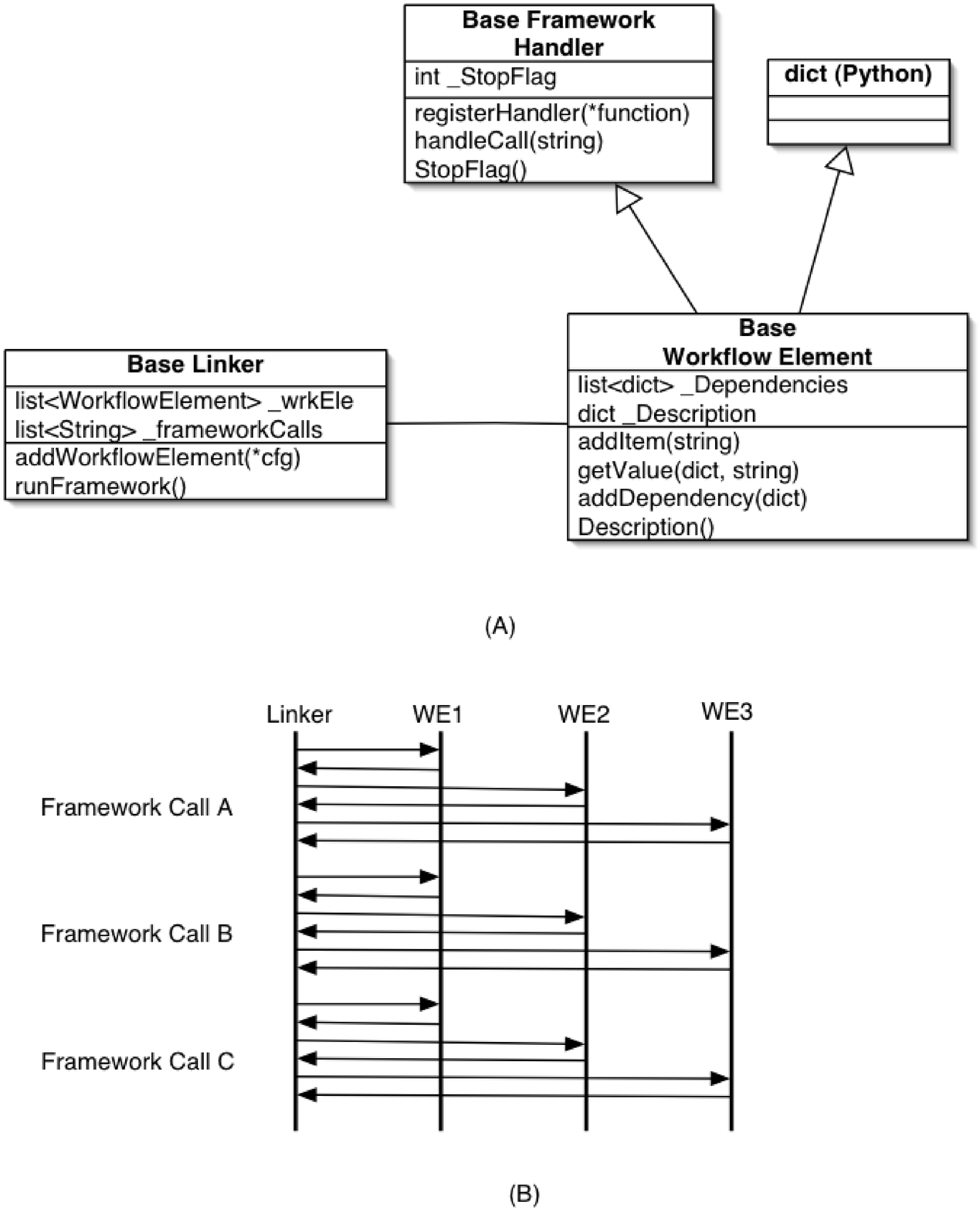}}
    \caption{(A) The simplified core class structure of RunJob.  Workflow elements 
consist of dictionaries that describe each application.  They in turn 
have descriptions consisting of key/value pairs and lists of dependencies
which reference descriptions of other workflow elements.  It can 
reduce a metadata flow using the GetValue method which references a metadata 
element in another workflow element by specifying elements from 
its description dictionary.  The workflow elements are constrained by an entity
called the Linker, which also issues framework messages to the workflow 
elements. Each workflow element may have a handler registered to handle the
framework message are task.  (B) An interaction diagram showing the pattern
of framework messages for three workflow elements (assumed to be in dependency 
order) and three framework messages or tasks.  Workflow elements may handle the framework 
call or not.}
    \label{fig:nfig2}
  \end{center}
\end{figure}

The RunJob Project was initiated at Fermilab as a common project between 
the DZero and the CMS experiments to combine their respective tools used in 
creating jobs for production of Monte Carlo data.  The software produced
contains the following elements:
\begin{itemize}
\item  Modular configuration.  Workflow elements consist of modular descriptions 
of application descriptions and dependencies upon other application descriptions.
\item  Context driven workflow.
\item  Simple framework model of workflow execution.
\item  Reduction algorithm for reducing metadata flows.
\end{itemize}
A simple diagram illustrating RunJob appears in figure \ref{fig:nfig2}.

\subsubsection{Modular Configuration}

Workflow elements in RunJob contain key/value parameters that 
configure the application at hand. These parameters can be 
used to run the application at hand directly or to create a job that 
will run the application later.  Workflow elements in RunJob are also 
given a complementary key/value pair description.  This description
is much like a Classed \cite{bib:classad} in Condor.  A given workflow 
element has dependencies expressed in terms of the descriptions of other 
workflow elements.  This is used to determine execution order of the workflow 
elements or of the jobs they create if they are making jobs.  

Finally, parameters in a workflow element are allowed to be 
references to parameters in other workflow elements by specifying 
the description of another workflow element to reference and the parameter 
within that workflow element to reference.  

\subsubsection{Context Driven Workflow}

In the RunJob software, a context document can be given that expresses simple 
constraints on the parameters known to the workflow elements.  
Each block of the context consists of the following information: 
\begin{itemize}
\item A header consisting of a workflow element description
\item A body consisting of a list of directives that can either 
constrain workflow parameters in individual workflow elements, 
define references between parameters of two different workflow 
elements, or redefine existing references between two different 
workflow elements.  
\end{itemize}
The context is loaded into the system prior to any workflow elements.
When a workflow element is added to the system, the directives within 
any context block whose header matches the description of the added 
workflow element are applied to that workflow element.  In the RunJob
software the match must be exact in the sense that all key/value elements
in the context block header must match the workflow element description.\footnote{This has been 
relaxed a bit in practice with wildcards.}  The workflow element description 
may have more key elements than the context block header in a match. 

Different context documents may be loaded into the system simultaneously. 
Currently, the RunJob system does not implement any conflict resolution 
algorithm on similar context blocks emanating from different documents.
Rather, the document specified last wins any conflict.

\subsubsection{Framework Model of Workflow}

In the RunJob software, the workflow is executed in framework fashion.  
The user can define a set of framework tasks, each framework task being 
represented by a string.  After all workflow elements have been added to 
the system, 
it will issue each framework task in the form of a message to each 
workflow element.  The order of the workflow elements is determined
by their dependencies.\footnote{In cases where there are no dependencies
to determine order, then the workflow elements appear in the order in which 
they were added to the system.}  The workflow elements may or may not 
respond to specific framework calls depending on how they are defined
and configured.

This arrangement is logically equivalent to a DAG in the case that 
there is one framework task and the nodes are laid out in dependency order.
The DAG has the advantage that in the case of a node failure, it may proceed 
to process independent nodes. A comparison of the methods of executing 
workflow elements is beyond the scope of this paper.  Many other workflow
systems do in fact use DAGMan, developed at the University of Wisconsin
at Madison, as a workflow execution manager.

\subsubsection{Reduction Algorithm}

Reduction in RunJob is accomplished by lazy evaluation of metadata flows.  
The metadata flow is evaluated by inter-workflow element lookup only when the target
of the metadata flow is accessed by some other entity or metadata flow.  This is 
accomplished by
setting read triggers on the attributes of RunJob workflow elements.

\subsection{MCRunjob}

MCRunjob \cite{bib:ggraham03} is a workflow management system based upon the 
RunJob software for the CMS experiment.  MCRunjob, however, is used for 
building jobs to produce Monte Carlo data; it does not actually execute
the jobs themselves.  Rather MCRunjob executes the workflow of "configuring
the jobs in the execution workflow."  This is nonetheless an ideal place to implement 
the ideas in this paper as they have to do with constraints on workflow 
configuration.  MCRunjob produces workflows for execution on 
DAGMan/Condor-G or on local batch systems. 

The simplified MCRunjob workflow in the example contains the following framework tasks.
\begin{itemize}
\item  contactDB: Contact the CMS Production Control Database
\item  configureJob: Reduce constraints by referencing metadata targets
\item  makeJob: Create and store application jobs into the Linker
\item  runJob: Retrieve and submit stored jobs
\end{itemize}
There is also an implicit reset framework task which is not shown.\footnote{
The details of how this workflow is executed efficiently to create N parallel jobs is 
unimportant here.  But briefly, there is a special group called the "onGroup" of 
framework calls that is executed N times.}  
If the workflow happens to contain a serialized DAG of
user applications, then the jobs are written out in the same serial order
and synchronized by some external means. 


\subsection{Practical Example}
\label{sc:example}

In this section, we present a simple workflow graph $G$
for the CMS experiment.  It is operated upon by a series 
of context documents $C_1$, $C_2$, and $C_3$ resulting 
in a final unconstrained workflow graph $M_G$ to be subsequently 
reduced by the RunJob lazy reduction algorithm.

The example here is based upon a real example from 
Monte Carlo production for the CMS experiment. 
It is not, however, an actual real life example.  The real life examples
contain far too many confounding specializations 
that are required in order to get actual Monte Carlo production done in the current 
imperfect but evolving Grid infrastructure.  Also, real CMS production
does not at this time constrain physics parameters by physics group.  
Rather, all actual production parameters come from a production control database
known as the RefDB \cite{bib:refdb}.  In one case, the 
syntax for the context block headers has been simplified.  Finally, 
many of the names of workflow elements and parameters have been changed 
to be more illustrative and less comprised of CMS jargon.  The applications 
described were discussed in section \ref{sc:applications}.

Consider the unconstrained workflow $G$ defined in section \ref{sc:uwrkflw}.  In 
this workflow, a user has chosen to run the generator program CMKIN 
followed by the simulation program OSCAR as well as the Digitization 
program.  The 
constraint that the input data of one step be read from the output of the 
previous step is expressed as a metadata flow.  
In MCRunjob, the workflow graph is just specifying the steps in the  
creation of a job, so a special "RunJob" workflow element is added at the end
to submit the jobs.  

The first context file $C_1$ is shown in section \ref{sc:frmwrk} and defines
the calls in the RunJob framework, described in section \ref{sc:applications}.
Calls are added to two predefined groups: "preGroup" which is executed only once, 
and "onGroup" which is executed as many times as there are jobs to be 
created.

The second context file $C_2$ is a hypothetical context file written by a 
physics group.  (The corresponding real file used for CMS Monte Carlo 
production is very long and unenlightening.)  It exhibits the for-each
and once-only semantics described in section \ref{sc:context}.  For 
each added workflow node matching a type declaration in the header of a
context block, metadata flows are added and dependencies are added.  
At the bottom of this 
context file, two metadata terminal nodes are added once only.  The RefDB
is a metadata source corresponding to the CMS production control database, 
and the PhysicsGroupDB is a metadata source corresponding to a 
hypothetical database that could be set up by
a physics group to organize parameters that people in the group should use. 

The third context file $C_3$ is a scheduler context file.  (The corresponding real
file used in CMS contains configurations for many different batch systems.) 
This context file substitutes a concrete choice of a workflow element,
LCG\_ResourceBroker,  for an abstract choice given in the workflow definition, 
RunJob. 

The context files given in the appendices do not collide on metadata targets
because they have been organized on the basis of the node types that they modify.
No collision algorithm is needed.  When the contexts are combined with the workflow, 
the final unconstrained 
workflow $M_G$ results.  It is given in section {sc:fuwrkflw}.  
As explained in section \ref{sc:applications}, 
RunJob reduces the unconstrained workflow into a fully constrained workflow
graph (not shown) using a lazy algorithm that triggers the evaluation
of a constraint when its target is accessed.  RunJob does not implement more
general Boolean constraints at this time.  

Note that if the user had to specify all of the constraints in the 
final unconstrained workflow, it would be a very complex task.  MCRunjob 
currently tracks about 150 parameters for CMS Monte Carlo production, 
mostly coming from the RefDB or describing
local site conditions.  As the system in CMS gets more complex and involves 
more applications and users, the number of parameters can only be expected 
to grow.  By splitting the work into different context files, 
the work can be shared among different roles: A naive user creates the 
initial unconstrained workflow in section \ref{sc:uwrkflw},
a developer maintains the framework context file in section \ref{sc:frmwrk}, 
a physics group 
convener maintains the physics context file in \ref{sc:physgroup}, 
and an administrator maintains the scheduler context file in section \ref{sc:sched}.

\section{Conclusion and Relation to Other Work}

This paper outlined some considerations for constraint modeling in Grid 
application workflows.  We focused mainly on semantic constraints and not 
constraints on physical resources or monitoring.  We have shown how 
multigraphs and metadata flows can be used to express constraints, 
and outlined a general procedure for reducing multigraphs into fully 
constrained graph workflow descriptions. Finally, we have shown how to 
factor out the constraints into contexts that can be maintained separately
and recombined later in a collaborative effort to constrain a workflow. 

\subsection{Related Work}

In addition to being used in creating jobs in MCRunjob, 
contexts have been demonstrated that govern the generation of fully 
constrained workflows as ordered lists of 
shell scripts, Condor-G/DAGMan \cite{bib:dagman}, 
or Chimera Virtual Data Language (VDL) \cite{bib:chimera} all from the 
same simple workflow.  The mechanism 
is to use a context to select one or more code generators. 
DAGMan is a complete workflow manager that has a very general model 
for specifying workflows, but it relies on the user completely to set up 
both the DAG and the parameters in the DAG itself.  VDL presents a 
unique view of data processing as a network of data products connected
by application induced transformations.  While it is possible for 
different collaborators to work independently on different transformations, 
there are not really tools for allowing collaborators to independently 
specify different aspects of the same chain of transformations.  The 
context mechanism presented here emphasizes rather the idea of virtual 
transformation as opposed to the virtual data.

Work on Context Oriented Programming \cite{bib:keays02} is being done 
for mobile computing.  Systems are being developed to exchange the actual 
code that gets run in different locales.  The present work is different in 
that it is effectively exchanging constraints and not actual code, although 
contexts can be fooled into loading different modules as shown in the example.  
Also, the definition of "locale" here is generalized to be any 
category of relevance to the VO: physics group, a personal role, et cetera.

Other workflow engines such as Triana \cite{bib:triana} and Webflow \cite{bib:webflow} 
emphasize graphical user interfaces and allow the user to visually determine
where individual elements are executed. Contexts are more suited to text based 
input. Since they are not graphs, they are more difficult to visualize.  

An interesting related work is GridAnt \cite{bib:gridant}.  The problem space of GridAnt is 
similar: it tries to tame the application space of the Grid by formulating 
the workflow as an Ant style Makefile with additional procedural constructs.
However, it too does not facilitate collaborative workflow in the same way 
that contexts do.

\subsection{Comparison to Cascading Style Sheets}

A fresh perspective on the context mechanism of RunJob may be obtained by 
considering Cascading Style Sheets (CSS)\cite{bib:css}.  A CSS document is loaded into a 
browser before processing HTML documents.  The CSS document consists of 
blocks with headers designed to match hierarchically defined segments of
an HTML document.  When an HTML document is loaded, each HTML element is 
processed according to configuration directives contained within a block 
of CSS corresponding to the segment in which the HTML element appears. 
A RunJob context is thus like CSS for workflow configuration.  Though the 
context block headers of RunJob don't appear to be hierarchical, they can
be if one chooses an ordered sequence of keys to use in workflow 
element descriptions. 

\subsection{Contextual Constraints and Provenance}

Provenance deals with the problem of collecting all of the information 
needed in order to recreate a data product.  The transformation graph 
approach of Chimera is most useful here \cite{bib:chimera}. 
Detailed provenance is therefore built into the system.  However, the 
prospect of saving metadata flows and metadata terminal nodes before the 
process of reduction begins on a constrained workflow graph offers the 
possibility of saving a new kind of provenance.  Namely, in addition to 
saving the flat values of all of the parameters that go into creating a 
data product, one can also save the constraints and relationships among 
those parameters and, by extension, why the parameters in a conventional 
provenance have the values that they do: which calibration set is being 
used, is it being used across all workflow steps, who signed off on the 
set of constraints as a whole and not just considering each constraint 
one by one.  This is because the provenance as expressed in a workflow 
constraint mechanism is categorical.

\appendix

\section{Example Workflow Specification Files}

The following sections contain code from CMS MCRunjob.  The language is 
called "macro language". Each statement in macro language is interpreted 
to specific calls on the RunJob API. In the future, the macro language 
will be deprecated in favor of using the RunJob API directly and expressing 
workflows and contexts in pure Python.  The statements of the macro 
language below are generally self explanatory.

\subsection{Unconstrained Workflow Definition}
\label{sc:uwrkflw}

This unconstrained workflow creates workflow elements for three CMS applications
described in section \ref{sc:applications} and links them together.  The 
condition that one application read input from the output of the previous 
step is expressed as a metadata flow.

\begin{verbatim}
attach CMKIN
attach OSCAR
OSCAR adddep CMKIN
OSCAR define inputFile ::CMKIN:outputFile
attach Digitization
Digitization adddep OSCAR
Digitization define inputDataset ::OSCAR:outputDataset
Digitization define inputRunNumber ::OSCAR:outputRunNumber
attach RunJob
framework run
\end{verbatim}

\subsection{Framework.ctx}
\label{sc:frmwrk}

The following context file simply defines the content and ordering of 
the framework calls issued by the Linker.  The "preGroup" is executed first
exactly once, and the "onGroup" is executed as many times as there are 
jobs to create.

\begin{verbatim}
framework define preGroup contactDB
framework define onGroup configureJob,makeJob,runJob
\end{verbatim}

\subsection{PhysicsGroup.ctx}
\label{sc:physgroup}

The following context file is a hypothetical physics group context.
It requires the addition of two metadata sources: the CMS production 
control database RefDB and a hypothetical physics group database.
Parameters for various applications are constrained either directly
in the context or are constrained by metadata flows from one or 
the other database.  Special directives are given to ensure that 
connections are opened when the Linker issues the "connectDB" framework
call.  The description keys are "Database" for the RefDB and the 
PhysicsGroupDB, and "Application" for the CMS physics applications.

\begin{verbatim}
contextBlock Database=PhysicsGroupDB,RefDB
    oncall contactDB do connectToDatabase
end
contextBlock Application=CMKIN,OSCAR,Digi
    add dependency Database=PhysicsGroupDB
    add dependency Database=RefDB
end
contextBlock Application=CMKIN 
  define ApplicationVersion 6.133 
  define ApplicationName kine\_make\_ntuple.exe 
  define HiggsMass ::PhysicsGroupDB:HMass2004
  define TopMass ::PhysicsGroupDB:TMass2004
end 
contextBlock Application=OSCAR 
  define ApplicationVersion OSCAR\_3\_6\_5 
  define HCal On
  define ECal On 
  define ECalThreshold :;PhysicsGroupDB:ECalThreshold2004
end
contextBlock Application=Digitization 
  define ApplicationVersion ORCA\_8\_4\_1 
  define PileupRate ::RefDB:Lumi_1032
end 
attach PhysicsGroupDB
attach RefDB
\end{verbatim}

\subsection{Scheduler.ctx}
\label{sc:sched}

The following context file defines the alias "RunJob" to mean
"LCG\_ResourceBroker".  When the novice user adds the generic 
"RunJob" element to a workflow, he/she gets the LCG\_ResourceBroker 
element when this context is loaded.  The context also contains
configuration information for the LCG\_ResourceBroker.  The 
tag "@args" means that the metadata flow comes from command
line arguments.

\begin{verbatim}
namespace add RunJob Scheduler=LCG_ResourceBroker
contextBlock Scheduler=LCG_ResourceBroker
  define UserJDLFile ::@args:UserJDLFile
  define ResourceBroker ::@args:ResourceBroker
  oncall RunJob do submit
end
\end{verbatim}

\subsection{Full unconstrained Workflow Definition}
\label{sc:fuwrkflw}

The above context files are loaded into the system first.  
(Remember that the PhysicsGroup.ctx also adds two 
metadata sources at this time.)  Upon addition of each element 
from the workflow of section \ref{sc:uwrkflw}, the 
metadata flows and other directives from matching 
context blocks are added to that element.  The following 
final unconstrained workflow appears below.  The 
metadata flows are reduced out as elements are accessed. 

\begin{verbatim}
framework define preGroup contactDB
framework define onGroup configureJob,makeJob,runJob
attach PhysicsGroupDB
PhysicsGroupDB oncall contactDB do connectToDatabase
attach RefDB
RefDB oncall contactDB do connectToDatabase
attach CMKIN
CMKIN add dependency Class=PhysicsGroupDB
CMKIN namespace add PhysicsGroupDB Class=PhysicsGroupDB
CMKIN add dependency Class=RefDB
CMKIN namespace add RefDB Class=RefDB
CMKIN define ApplicationVersion 6.133 
CMKIN define ApplicationName kine\_make\_ntuple.exe 
CMKIN define HiggsMass ::PhysicsGroupDB:HMass2004
CMKIN define TopMass ::PhysicsGroupDB:TMass2004
attach OSCAR
OSCAR add dependency Class=PhysicsGroupDB
OSCAR namespace add PhysicsGroupDB Class=PhysicsGroupDB
OSCAR add dependency Class=RefDB
OSCAR namespace add RefDB Class=RefDB
OSCAR define ApplicationVersion OSCAR\_3\_6\_5 
OSCAR define HCal On
OSCAR define ECal On 
OSCAR define ECalThreshold :;PhysicsGroupDB:ECalThreshold2004
OSCAR adddep CMKIN
OSCAR define inputFile ::CMKIN:outputFile
attach Digitization
Digitization add dependency Class=PhysicsGroupDB
Digitization namespace add PhysicsGroupDB Class=PhysicsGroupDB
Digitization add dependency Class=RefDB
Digitization namespace add RefDB Class=RefDB
Digitization define ApplicationVersion ORCA\_8\_4\_1 
Digitization define PileupRate ::RefDB:Lumi_1032
Digitization adddep OSCAR
Digitization define inputDataset ::OSCAR:outputDataset
Digitization define inputRunNumber ::OSCAR:outputRunNumber
attach LCG_ResourceBroker
LCG_ResourceBroker define UserJDLFile ::@args:UserJDLFile
LCG_ResourceBroker define ResourceBroker ::@args:ResourceBroker
LCG_ResourceBroker oncall RunJob do submit
framework run
\end{verbatim}


\begin{thebibliography}{20}


\bibitem{bib:foster04} I. Foster and C. Kesselman, ed. The Grid: Blueprint for a New Computing Infrastructure, 2nd Edition, Morgan Kaufmann (2004)

\bibitem{bib:berman03} F. Berman, G. Fox, and T. Hey. Grid Computing: Making The Global Infrastructure a Reality, John Wiley \& Sons, (2003)

\bibitem{bib:grid2003} Grid2003 http://www.ivdgl.org/grid2003

\bibitem{bib:egee} Enabling Grids for E-sciencE http://public.eu-egee.org/

\bibitem{bib:osg} Open Science Grid http://www.opensciencegrid.org

\bibitem{bib:mpi} Gropp, W. et al.  Using MPI: Portable Parallel Programming with the Message-Passing Interface. The MIT Press, Cambridge, Massachusetts, 3rd printing,  1996.

\bibitem{bib:runjob} The RunJob Project homepage http://projects.fnal.gov/runjob
\bibitem{bib:ggraham03} G.E. Graham, et al. MCRunjob: A High Energy Physics Workflow Planner for Grid Production Processing Proceesings of CHEP 2003 (TUCT007) San Diego

\bibitem{bib:cms} The CMS Experiment homepage: http://cmsinfo.cern.ch


\bibitem{bib:cern} CERN Laboratory homepage: http://www.cern.ch

\bibitem{bib:dagman} DAGMan: http://www.cs.wisc.edu/condor/dagman

\bibitem{bib:classad} Raman, R. and Livny, M.  "Matchmaking: Distributed Resource 
Management for High Throughput Computing"  Proceedings of the Seventh IEEE International 
Symposium on High Performance Distributed Computing,  July 28-31, 1998, Chicago, IL.

\bibitem{bib:chimera} I. Foster, et al. "Chimera: A Virtual Data System for Representing, Querying, 
and Automating Data Derivation"  14th International Conference on Scientific and Statistical Database 
Management (SSDBM 2002)

\bibitem{bib:pierce91} B. Pierce.  Basic Category Theory for Computer Scientists.  MIT Press (1991) 


\bibitem{bib:atlas} The ATLAS Experiment homepage: http://atlas.web.cern.ch/Atlas/

\bibitem{bib:refdb} CMS Internal Note 2002/044  http://cmsdoc.cern.ch/cms/
                                   production/www/documents/RefDB/in02\_044.ps

\bibitem{bib:keays02} R. Keays. Context Oriented Programming.  PhD Thesis, University of Queensland, Australia (2002)
 

\bibitem{bib:triana} "Triana Workflow" http://www.triana.co.uk

\bibitem{bib:webflow} Bhatia, D. et al.  "WebFlow - A Visual Programming Environment for Web/Java
Based Coarse Grain Distributed Computing," Concurrency: Practice and Experience, vol. 9, no. 6, p. 555
(1997)

\bibitem{bib:gridant} von Laszewski, G. et al.  "GridAnt: A Client-Controllable Grid Workflow System"
37th Hawaii International Conference on Computer Science, Hawaii, USA (2004)

\bibitem{bib:css} Cascading Style Sheets  http://www.w3.org/Style/CSS/


\end{thebibliography}
\end{document}